\documentclass[11pt,psfig]{article} \textheight=23 true cm \textwidth=16 true cm
\usepackage[dvips]{graphicx}            
\usepackage{pslatex}	    
\usepackage{subfigure}
\usepackage{cite}
\usepackage{graphicx}
\usepackage{amsmath}
\usepackage{amssymb}
\usepackage{color}
\usepackage{mathtools}
\begin{document}
\begin{center}
{\Large\bf Accelerating Universe in Higher Dimensional Space Time : an alternative approach }\\[8 mm]
D. Panigrahi\footnote{ Sree Chaitanya College, Habra 743268, India, \emph{and also} Relativity and Cosmology Research Centre, Jadavpur
University, Kolkata - 700032, India , e-mail:
dibyendupanigrahi@yahoo.co.in },
 B. C. Paul\footnote{Department of Physics, University of North Bengal, Dist.-Darjeeling, PIN-734013, India, e-mail : bcpaul@associates.iucaa.in}
  and S. Chatterjee\footnote{ New Alipore College (retd.), Kolkata - 700053, India, \emph{and also} Relativity and Cosmology Research Centre, Jadavpur
University, Kolkata - 700032, India , e-mail : chat\_sujit1@yahoo.com} \\[10mm]

\end{center}

\begin{abstract}
We have discussed here a higher dimensional cosmological model and
explained the recent acceleration with a Chaplygin type
of gas.   Dimensional reduction of extra space is possible in this case.
 Our solutions are general in nature because all the well
known results of 4D Chaplygin driven cosmology are recovered when $d = 0$.
 We have drawn the best fit graph using  the data obtained by the differential
 age method (CC) and it is seen that the graph favours only one extra dimension.
  That means the Chaplygin gas is apparently dominated by a 5D world.
   Relevant to point out that the final equation in this case are highly nonlinear in nature.
    Naturally it is not possible to obtain explicit solution of the 4D scale factor with time.
    To circumvent this difficulty, we consider a first order approximation of the key equation
    which has made it possible to get time explicit solution of 4D scale factor in
    exact form as well as the expression of extra dimensions. It may be pointed out that for large four dimensional scale factor this solution mimics $\Lambda$CDM model.     An analysis of flip time is also studied both analytically and graphically in some detail. It  clearly shows that early \emph{flip} occurs for higher dimensions. It is also seen that the rate of dimensional reduction is faster for higher dimensions. So we may conclude that the effect of compactification of extra dimension  helps the acceleration.

\end{abstract}

\bigskip

\bigskip
\section {Introduction :}

There has been a resurgence of interests in models where the present universe  seems to be undergoing
an accelerated expansion. Gravitational force being always attractive in nature, this finding is contrary to our intuition.
However, detail investigations of redshifts of type Ia distant nabulae as well as cosmic
microwave background anisotropy measurements did suggest the accelerated type of  expansion.
Several explanations present themselves - introduction of higher derivative theories ~\cite{ua} , a variable cosmological constant
in Einstein's field equation ~\cite{nas}, flavor
oscillations of axions ~\cite{kalop}, inhomogeneity in space time structure ~\cite{aln, dp0},
quintessential type of scalar
 field ~\cite{pd}, presence of higher dimensions ~\cite{dp1}, and most importantly a Chaplygin type of gas
 as a matter field~\cite{kp}.

 In this context the authors of the present
article have been, of late, struggling
 with the idea of explaining the late acceleration as a higher
 dimensional(HD) phenomena ~\cite{dp, sc1}. In the framework of higher dimensional
 cosmology we have been able to show, though in a rather \emph{naive}
 way, that the acceleration can be explained as a consequence of
 the presence of the extra spatial dimensions and this effect has
 been  coined as `\emph{dimension driven}' accelerating model.
 In fact here the effective Friedmann equations contain additional
 terms resulting from the presence of  extra dimensions which may be  interpreted as a
 `fluid' causing  the late acceleration. So in this work we
 attempt to incorporate the phenomenon of acceleration within the
 framework of higher dimensional spacetime itself without invoking a mysterious
 scalar field with large negative pressure by hand. Moreover,
 the origin of the extra fluid responsible for the acceleration is
 geometric in origin having strong physical foundation and more in
 line with the spirit of general relativity as proposed by Einstein~\cite{einstein}
 and later developed by Wesson and his collaborators ~\cite{wes}. In an earlier
 work Milton ~\cite{milton} has shown  that quantum fluctuations in
 4D spacetime do not generate the dark energy but rather a
 possible source of the dark energy is the fluctuations in the
 quantum fields including quantum gravity inhabiting extra
 compactified dimensions. This has led a number of workers to concentrate on ideas
  relating to higher dimensional space in its
 attempts to unify gravity with other forces of nature,
 interpretation of different brane models, space-time-matter (STM) proposal ~\cite{wes} and
also the dimension driven quintessential models ~\cite{pie}.
 The present investigation is primarily motivated by two considerations.
 While we have plenty of multidimensional cosmological models in
 literature~\cite{bron} and also some sporadic works of  brane models ~\cite{mak, her}
 with Chaplygin type of fluid, but scant attention has been paid so far to explain the cosmic acceleration either by extra dimensions themselves or by Chaplygin type of matter field ~\cite{ran, sal}.

\vspace{0.01cm}

 The present work essentially contains two parts.
We have taken a
 (d+4) dimensional homogeneous spacetime with two scale factors and a
 perfect fluid as a source field. Here we have taken
  a Chaplygin type of matter field  with higher dimensional spacetime.
  The solution of our key equation \eqref{eq:12}  in closed form cannot be obtained
  because integration yields an elliptical solution only and we get
   hypergeometric series.  In any case certain inferences can always be drawn in the
 extreme cases and our analysis shows that an initially
 decelerating model transits to an accelerating one as in 4D. An
 interesting result in this section is the fact that the effective
 equation of state(EOS) at the late stage of evolution contains some
 additional terms coming from extra dimensions. This finding has
 marked similarity with the EOS obtained by Guo \emph{et al. }~\cite{guo} for a variable
  Chaplygin gas model. Depending upon the presence of extra dimensions the
  cosmology then evolves as  $\Lambda$CDM or Phantom
 type. This is definitely at variance with the usual 4D models which
 essentially ends up in a deSitter phase with time.   Though not
 exactly similar this points to the `\emph{k-essence}' type of models
 which lead to cosmic acceleration today for a wide range of initial
 conditions without fine-tuning and without invoking an anthropic
 argument.

 We adopt here $\chi^2$ minimization technique
to obtain constraints imposed by cosmological observations. We use Type Ia Supernova
data and the predictions of CMB, BAO in constraining the cosmological models.
Defining a total $\chi^2$ function, we analyse
cosmological models using the $(H(z) - z)$ OHD data (table 1).
The constraints on the $\Omega_m$ and $m$ (to account for dimensional reduction $m > 0$) are determined by
drawing contour plots at different confidence levels. We have drawn the
best fit graph using  the data obtained by the
differential age method (CC) and it is seen that the graph favours only one extra
dimension. That means the Chaplygin gas  apparently mimics  a 5D world.

It is to be mentioned here that dimensional reduction of extra space is possible in this case.
 But we can not explain the impact of compactification of extra dimensions on
 present acceleration  or the evolution of scale factor of the universe
 because the key eq. \eqref{eq:12} is not amenable to obtain an explicit
 solution, so we have to study the extremal cases only.
 This type of incompleteness may be remedied via an \emph{alternative approach} ~\cite{dp2}
 where the higher order terms of the binomial expansion of RHS  of eq. \eqref{eq:12}
   are neglected. The main reason behind it is that the 4D scale factor
   should be large enough at zero pressure era and it may not be inappropriate if
   we take only the first order terms of the binomial expansion of RHS of the  eq. \eqref{eq:12}
    which was shown in eq. \eqref{eq:22}. In the process we have got an exact solution
    through which we study an explicit time dependent solution.

\section {Higher Dimensional Field Equations :}

The Einstein field equation in $d-$ dimension is given by
\begin{equation}\label{eq:0}
R_{AB} - \frac{1}{2} g_{AB} R = \kappa T_{AB}
\end{equation}
where $A, B$ are $(0,1,2,3, .....d)$, $R_{AB}$ and $R$ are the Ricci tensor and Ricci scalar respectively.
We  consider the line element of (d+4)-dimensional spacetime
\begin{eqnarray}\label{eq:1}
  ds^{2} &=&
  dt^{2}-a^{2}(t)\left(\frac{dr^{2}}{1-r^{2}}+r^{2}d\theta^{2}+
  r^{2}sin^{2}\theta d\phi^{2}\right) - b^{2} (t)\gamma_{\alpha \beta}dy^{\alpha}dy^{\beta}
\end{eqnarray}
where $y^{\alpha}$ ($\alpha, \beta = 4,....        , 3+d$) are the extra
dimensional spatial coordinates and the 3D and extra dimensional
scale factors $a(t) $ and $b(t) $ depend on time only  and the compact manifold is described by the metric
$\gamma_{ab}$. We consider the manifold $M^{1}\times S^{3}\times S^{d}$
the symmetry group of the spatial section is $O(4) \times O(d+1)
$. The stress tensor whose form will be dictated by Einstein's
equations must have the same invariance leading to the energy
momentum tensor as \cite{rd}
\begin{equation}\label{eq:2}
T_{00}=\rho~,~~T_{ij}= -p(t)g_{ij}~,~~ T_{\alpha \beta}=-p_{d}(t)g_{\alpha \beta}
\end{equation}
where the rest of the components vanish. Here $p$ is the isotropic
3-pressure and $p_{d}$, that in the extra dimensions.
\vspace{0.5cm}
Considering,
\begin{equation}\label{eq:3}
b(t) = a(t)^{-m}
\end{equation}
where $m$ is any positive number so that
dimensional reduction is ensured  \emph{a priori}. For the matter
field we here assume an equation of state given by the
Chaplygin type of gas in 3D space only ~\cite{kam} which is
\begin{equation}\label{eq:4}
p= - \frac{B}{ \rho}
\end{equation}
The field equations are given by \cite{dp}
\begin{eqnarray}
\rho = \frac{k}{2} \frac{\dot{a}^{2}}{a^{2}}\label{eq:5} \label{eq:5}~~~~~~~~~~~~~~~~~~~~~~~~~~~~~~~~~~~~~~~~~~~~~~~~~~~\\
 -p =
(2-md)\frac{\ddot{a}}{a} + \frac{1}{2}[m^{2}d( d+1) +
2(1-md)]\frac{\dot{a}^{2}}{a^{2}} \label{eq:6} \\
-p_{d}= (3-md +
m)\frac{\ddot{a}}{a}+\frac{1}{2}[m(d-1)(md-4)+6]\frac{\dot{a}^{2}}{a^{2}} \label{eq:7}
\end{eqnarray}
where $k = m^{2} d(d-1) + 6(1-md)$. For a positive energy density, $k$ must be greater
than zero which implies $m < \frac{3d-\sqrt{3d(d+2)}}{d(d-1)}$
 or, $m > \frac{3d+\sqrt{3d(d+2)}}{d(d-1)}$.

\begin{figure}[ht]
\centering \subfigure[ Figure shows the  $d \geq 0$ ]{
\includegraphics[width= 6.8 cm]{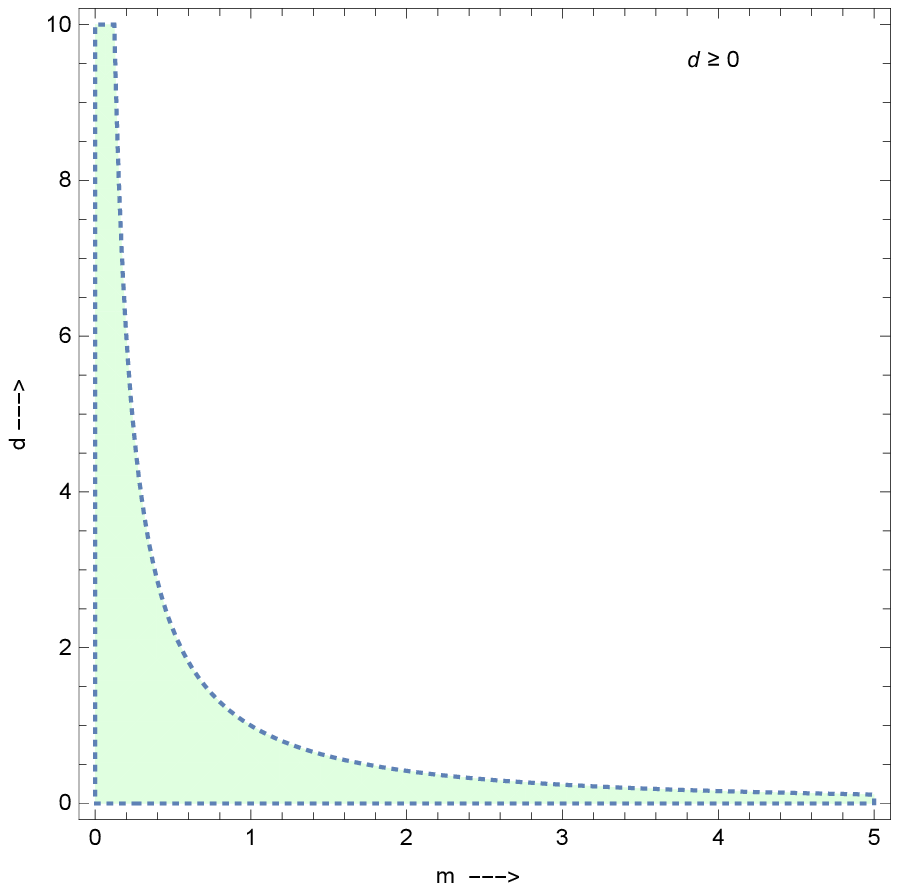}
\label{fig:subfig1} } ~~~\subfigure[  Figure shows the $m \leq 1  $ for $d \geq 1$  .]
{
\includegraphics[width= 6.8 cm]{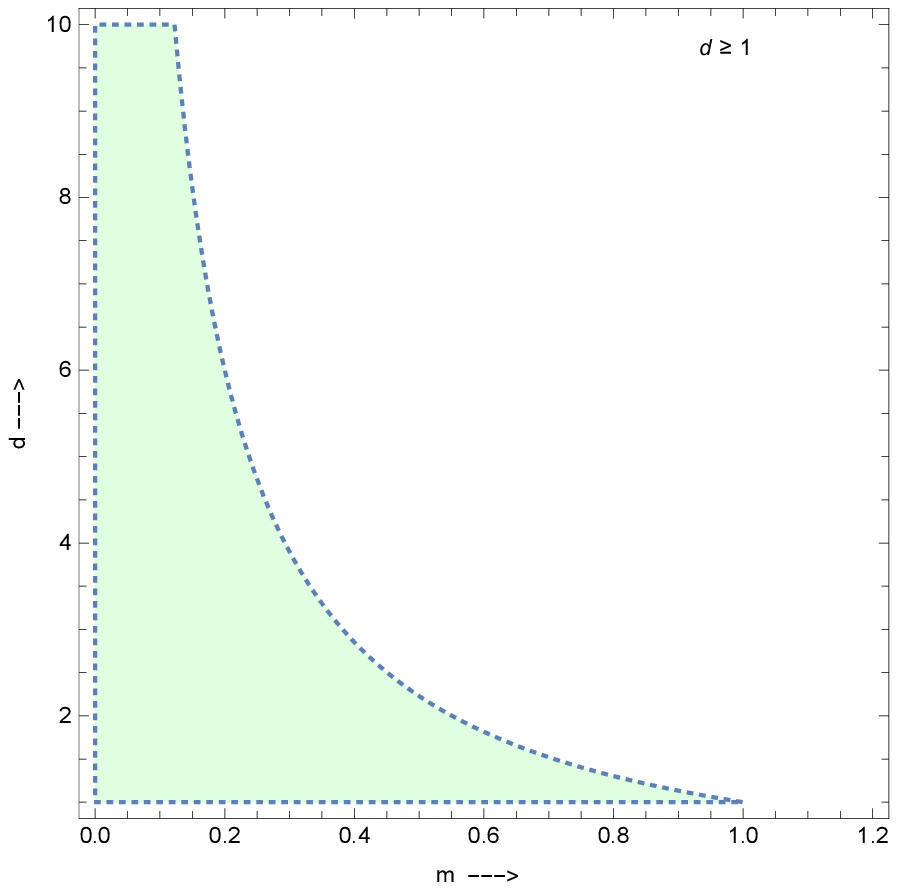}

 \label{fig:subfig2} }\label{fig:subfigureExample}~~~~~~~~~~~\caption[Optional
caption for list of figures]{\emph{ The region plot  of $d$ vs $m$ is shown in this figure with the conditions $`k > 0$' and $`md <2$'.  }}
\end{figure}
\vspace{0.5 cm}
The conservation equation is given by
\begin{equation}\label{eq:8}
\dot{\rho} + 3(\rho + p)\frac{\dot{a}}{a} +
d \;(\rho + p_{d})\frac{\dot{b}}{b} = 0
\end{equation}

Now using eqs. \eqref{eq:6}, \eqref{eq:7} \& \eqref{eq:8} we get
\begin{eqnarray}\label{eq:9}
\dot{\rho} + \frac{k}{(2-dm)}\frac{\dot{a}}{a}\left[\left\{1
+ \frac{2 dm (m+1)}{k}\right\}\rho - B\rho^{-1}\right] =0
\end{eqnarray}
Solving eq. \eqref{eq:9} we get
\begin{eqnarray}\label{eq:10}
\rho = \left[ \frac{Bk}{M} +
\frac{c}{a^{\frac{2 M}{(2-dm)}}}\right]^{\frac{1}{2}}
\end{eqnarray}
where
\begin{equation}\label{eq:11}
M = k + 2 d \; m (m+1)
\end{equation}
 and
$c$ is the integration constant. From physical considerations we determine the restriction on $m$ as $m <
\frac{3d-\sqrt{3d(d+2)}}{d(d-1)}$ for $ d \neq 1$ otherwise $m <1$ for $
d = 1$. It may be mentioned here that a  detail analysis was given by two of us in Ref. \cite{dp}
 in a  Modified Chaplygin gas cosmology.
With the constraint \eqref{eq:11}, the
last term in the right hand side is obtained from higher dimensional contribution which is absent in  4D
($d=0$). Thus the density of the universe at the present epoch in the framework of a higher dimensional is less compared to a 4-dimensional universe.  We note the following :

(\emph{i}) when  $m = - 1$ we get a universe with $ b (t) = a (t)$, which permits a universe
with  expansion in all dimensions. In this case the observed universe with
desirable feature of dimensional reduction to get an isotropic
expansion is not obtained.

 (\emph{ii}) when  $m = 0$, a universe with flat extra space in $(d+ 3)$ dimensions is obtained.  In this case also we get
 similar scenario that is obtained in a 4D universe which was reported in  Ref. ~\cite{ud}.
In fact
 the similarity is a direct consequence of a  known theorem of Campbell
 that any analytic N-dimensional Riemmanian manifold can be
 locally embedded in a higher dimensional Ricci-flat manifold
 ~\cite{tavako}.

(\emph{iii})  when $d=0$, we  recover the 4D metric with all
the known features of  4D cosmology.\vspace{0.5cm}

Using eqs. \eqref{eq:5}  \&  \eqref{eq:10}, we get
\begin{eqnarray}\label{eq:12}
\frac{\dot{a}^{2}}{a^{2}} = \frac{2}{k} \left[ \frac{Bk}{M} +
\frac{c}{a^{\frac{2 M}{(2-dm)}}}\right]^{\frac{1}{2}}
\end{eqnarray}
The known form of the scale factor can be obtained from the above equation.
The solution of the eq. \eqref{eq:12} in closed form cannot be obtained because integration yields an elliptical solution only and we get hypergeometric series. However the eq. \eqref{eq:12}  gives significant information under extremal conditions as briefly discussed here.

\section { Cosmological dynamics : }
Now we have discussed the cosmological behaviour of the Chaplygin gas equation of state in higher dimensional spacetime.
We have expressed the relevant equations with the help of deceleration parameter. In what follows we shall see from the observational data, the best fit graph favours a 5-dimensional interpretation of the cosmological dynamics.

\vspace{0.5 cm}
\subsection{Deceleration Parameter:}

At the early stage of the cosmological evolution
when the scale factor $a(t)$ of the universe is relatively small, the second term
of the right hand side of the eq. \eqref{eq:12} dominates which has been
discussed in the literature ~\cite{bento}. Using the expression of the deceleration parameter,
$q$ we get
\begin{equation}\label{eq:13}
q = -\frac{1}{H^{2}}\frac{\ddot{a}}{a}= \frac{d}{dt} \left(H^{-1}
\right) - 1 =  -1 + \frac{M}{2(2-md)} + \frac{k}{2(2-md)} \frac{p}{\rho}
\end{equation}
where $H$ is the Hubble constant. With the help of the equation of state (EoS) given
by \eqref{eq:4} we get
\begin{equation}\label{eq:14}
q =  -1 + \frac{M}{2(2-md)} - \frac{kB}{2(2-md)} \frac{1}{\rho^2}
\end{equation}
one again using eq. \eqref{eq:10} we obtain
\begin{equation}\label{eq:15}
q = -1 + \frac{M}{2(2-md)} -\frac{Bk}{2(2-md)}\left(\frac{Bk}{M}+\frac{c}{a^{\frac{2M}{2-md}}}\right)^{-1}
\end{equation}
Again at flip time, {\it i.e.} when $q = 0$ the scale factor becomes
\begin{equation}\label{eq:16}
a_{flip} = \left[\frac{c}{\frac{kB}{M-2(2-dm)}-\frac{Bk}{M}}\right]^{\frac{2-dm}{M}}
\end{equation}
where $a_{flip}$ signifies the sign change of the deceleration parameter.
Again, in terms of redshift parameter $\left(1+z = \frac{a_0}{a}\right)$
 where $a_0$ is the scale factor of the present universe,  we can re-write
  the eq. \eqref{eq:15} as
\begin{equation}\label{eq:17}
q = -1 + \frac{M}{2(2-dm)} -\frac{Bk}{2(2-dm)}\left[\frac{Bk}{M}+ c \left(\frac{1+z}{a_{0}}\right)^{\frac{2M}{2-dm}}\right]^{-1}
\end{equation}
and the redshift parameter at the flip epoch ($z_f$) is given by
\begin{equation}\label{eq:18}
z_f = \left[\frac{1}{c} \left(\frac{Bk}{M-2(2-dm)}- \frac{Bk}{M}\right) \right]^{\frac{2-dm}{2M}}a_{0} -1
\end{equation}
As the universe expands the energy density  $\rho$ decreases with time such that the
last term in the eq. \eqref{eq:14} increases indicating  a sign flip when the density attains a critical value given by
\begin{equation}\label{eq:18a}
\rho = \rho_{flip} = \left[ \frac{Bk}{M-2(2-md)}\right]^{\frac{1}{2}}
\end{equation}
It is evident that for $M > 2(2-dm)$  one gets a universe with normal matter.  This is a consistent result for a realistic $z_{f}$.

In the next section we discuss  the extremal cases to understand the evolution of the universe. Similar cases in 4 dimensional universe with modified Chaplygin gas is discussed in \cite{grav, pro}.
\vspace{0.5 cm}

\textbf{CASE~A :}~~ In the early phase when the scale factor
$a(t)$ is very small, the  eq. \eqref{eq:15} reduces to
\begin{equation}\label{eq:19}
q = -1 + \frac{M}{2(2-dm)}\\
=\frac{1}{2} + \frac{dm(m+1)}{2(2-dm)} - \frac{dm}{2}
\end{equation}
representing a dust dominated universe. It is found that $q = \frac{1}{2}$ for $d = 0$, {\it i.e.,} in a 4-dimensional space time, which is in consonance with well known 4D results.

\vspace{0.5 cm}

 \textbf{CASE~B :}
 In the later epoch  of evolution, {\it i.e.}, for a large size of the universe,  we get from the eq. \eqref{eq:15}
\begin{equation}\label{eq:20}
q = -1
\end{equation}
which is similar to that one finds in a $\Lambda$CDM model. It further gives the \emph{effective} EoS using the eq. \eqref{eq:13},
\begin{equation}\label{eq:21}
\mathcal{W}  =  \frac{p}{\rho} =  - \left[ 1 + \frac{2d\, m(m+1)}{k} \right]
\end{equation}
It is interesting to note that the \emph{effective} EoS
is not time dependent. In what follows we shall find that at the later stage of
evolution of the universe as $a (t) \rightarrow \infty$, $\mathcal{W}
\rightarrow -1 $ so in the asymptotic region it can be expressed as  $ p = - \rho$ even one begins with a modified Chaplygin gas which corresponds to an empty universe. It corresponds to a
universe with a cosmological constant from eq. \eqref{eq:20} it is evident
that the deceleration parameter, $q $ reduces to $-1$. Again in the presence of
extra dimensions $( d \neq 0 )$, eq. \eqref{eq:21}  points to a phantom type
 $ (\mathcal{W} < -1)$ but in the analogous 4D case $(d = 0)$, it mimics a $\Lambda$CDM model.
 This striking difference results from the appearance of extra terms coming from additional dimensions in EoS.

\vspace{0.5 cm}
\subsection{Observational Constraints on  the  Model Parameters:}
\vspace{0.3 cm}

In this section the  observational data ~\cite{pur} will be used to  analyze cosmological model estimating the constraints imposed on the model parameters. We use Type Ia Supernova data and the predictions of CMB, BAO in constraining the cosmological models.
It is difficult to integrate the expression to determine the exact temporal behaviour of the scale factor we use an alternative way to express the expansion rate of the universe as a function of redshift, {\it i.e.}, $ H(z)$ ~\cite{sei} in the data analysis.
 In  our case the observed Hubble data (OHD)  set,   the most direct and model independent observable of the dynamics of the universe will be used here in the model. Naturally, the $H(z)$ dataset here shows the fine structure of the expansion history of the universe.
One can not get the Hubble $H(z)$ data  directly  from
a tailored telescope. Instead, one may get it from two different
methods. First is to calculate the differential ages of galaxies ~\cite{sim,ster}, usually called cosmic chronometer (CC),
other  is to the deduction from the radial BAO peaks in
the galaxy power spectrum ~\cite{gaz, mor}  or from the BAO peak using the $Ly-\alpha$
forest of QSOs ~\cite{del}  based on the clustering of galaxies or quasars. We analyze the cosmological model using the
compilation of OHD data points collected by Magana \emph{et al.} ~\cite{mag}
 and Geng \emph{et al.} ~\cite{gen},  the  $H(z)$
data reported in various surveys so far. The $31 $ CC $ H(z)$
data points are listed in Table 1.

\vspace{0.5 cm}
\begin{table}[h!]
\small
  \centering
  \caption{The latest Hubble parameter measurements $H(z)$ (in units of $km s^{-1}Mpc^{-1}$ ) and their errors $\sigma$  at redshift $z$ obtained from the differential age method (CC). }\vspace{0.5 cm}
  \label{tab:table1}
\begin{tabular}{|c|c|c|c|}  \hline
  z & H(z) & $\sigma$ &  Reference \\   \hline \hline
  0.0700 & 69.00 & $\mp 19.6 $& \ ~\cite{zhang}\\
  0.1000 & 69.60 & $\mp 12.00 $ &   ~\cite{ster}\\
  0.1200 & 68.60 & $\mp 26.20$ &  ~\cite{zhang}\\
  0.1700 & 83.00 & $\mp 8.00 $&   ~\cite{ster}\\
  0.1797 & 75.00 & $ \mp 4.00$ &  ~\cite{mor1}\\
  0.1993 & 75.00 & $\mp 5.00$ &  ~\cite{mor1}\\
  0.2000 & 72.90 & $\mp 29.60 $&  ~\cite{zhang} \\
  0.2700 & 77.00 &$ \mp 14.00$ &  ~\cite{ster}\\
  0.2800 & 88.80 & $\mp 36.60 $& ~\cite{zhang}\\
  0.3519 & 83.10 & $\mp 14.00$ &  ~\cite{mor1}\\
  0.3802 & 83.45 & $\mp 13.50 $&  ~\cite{mor2}\\
  0.4000 & 95.00 &$ \mp 17.00$ &   ~\cite{ster}\\
  0.4004 & 77.00 &$ \mp 10.20 $&   ~\cite{mor2} \\
  0.4247 & 87.10 & $\mp 11.20 $&   ~\cite{mor2}\\
  0.4497 & 92.80 &$ \mp 12.90 $&  ~\cite{mor2} \\
  0.4700 & 89.00 &$ \mp 34.00 $&  ~\cite{rats}\\
  0.4783 & 80.90 &$\mp 9.00 $&   ~\cite{mor2}\\
  0.4800 & 97.00 & $\mp 60.00$ &  ~\cite{ster}\\
  0.5929 & 104.00 &$ \mp 13.00 $&  ~\cite{mor1}\\
  0.6797 & 92.00 & $\mp 8.00$ &  ~\cite{mor1}\\
  0.7812 & 105.00 & $\mp 12.00$ &  ~\cite{mor1}\\
  0.8754 & 125.00 & $\mp 17.00$ & ~\cite{mor1} \\
  0.8800 & 90.00 & $\mp 40.00 $ & ~\cite{ster} \\
  0.9000 & 117.00 &$ \mp 23.00 $&   ~\cite{ster}\\
  1.0370 & 154.00 &$\mp 20.00 $&   ~\cite{mor1}\\
  1.3000 & 168.00 & $\mp 17.00$ &  ~\cite{ster}\\
  1.3630 & 160.00 &$ \mp 33.60 $&  ~\cite{mor3}\\
  1.4300 & 177.00 & $\mp 18.00$ &   ~\cite{ster}\\
  1.5300 & 140.00 & $\mp 14.00$ & ~\cite{ster}\\
  1.7500 & 202.00 & $\mp 40.00$ &  ~\cite{ster} \\
  1.9650 & 186.50 & $\mp 50.40 $ &  ~\cite{mor3} \\
   \hline
\end{tabular}
\end{table}
     The Hubble parameter depending on the differential ages as a function of
     redshift $ z$ can be written in the form of

\begin{equation}\label{eq:20a}
H(z) = -\frac{1}{1+z} \frac{dz}{dt}
\end{equation}
therefore, from eq. \eqref{eq:20a}  $H(z)$ can be found directly once $\frac{dz}{dt}$ is known ~\cite{sei}.
We consider the Hubble parameter $H$ and the three space scale factor $a$, then the eq. \eqref{eq:5} may be expressed as $ \rho =\frac{k}{2} H^2$. Using the present value of the scale factor  normalised to unity , \emph{i.e.}, $a = a_0 =1$ we get a relation of the Hubble parameter with the redshift parameter $z$. If $\rho_0$ be the density at present epoch  then the well known density parameter can be written as   $\Omega_m = \frac{c}{\frac{BK}{M} + c}$ ~\cite{seth}. Now using eq. \eqref{eq:10}, we can express the three space matter density as
\begin{equation}\label{eq:21a}
\rho = \rho_0 \left \{1-\Omega_m + \Omega_m \left(1+z \right)^{\frac{2M}{2 -dm}} \right \}^{\frac{1}{2}}
\end{equation}
and the Hubble parameter
\begin{equation}\label{eq:21b}
 H(z) = H_0 \left \{1-\Omega_m + \Omega_m \left(1+z \right)^{\frac{2M}{2 -dm}} \right \}^{\frac{1}{2}}
\end{equation}
where $H_0 = \left(\frac{2 \rho_0}{k} \right)^{\frac{1}{2}}$ is the present value of the Hubble parameter.
The eq. \eqref{eq:21b} shows the evolution of Hubble parameter $H (z)$ as a function of redshift parameter $z$. The graphical presentation of \eqref{eq:21b} is shown in Fig-$2(b)$ where it has been compared with best fit curve.
We draw a best fit curve  of redshift against Hubble parameter in the $1 \sigma$ confidence region from the data given by the Table-1. The dots in figure-$2(a)$ indicate the recent observable values. Here  the value of $H_0 = 70 ~ Km S^{-} Mpc^{-1}$ with $\Omega_m = 0.18$ is taken.
In the figure-$2(b)$ we  compared the   best fit curve with the theoretical expression of $H(z)$ obtained in eq. \eqref{eq:21b}. It is evident  that for $d = 1$ graph almost coincides with the best fit graph for $m = 0.54$. Thus it gives hint of a universe with one extra dimension which is  more acceptable with $m=0.54$.

\begin{figure}[ht]
\centering \subfigure[ Figure shows the 'Best fit curve' from Planck data ]{
\includegraphics[width= 6.8 cm]{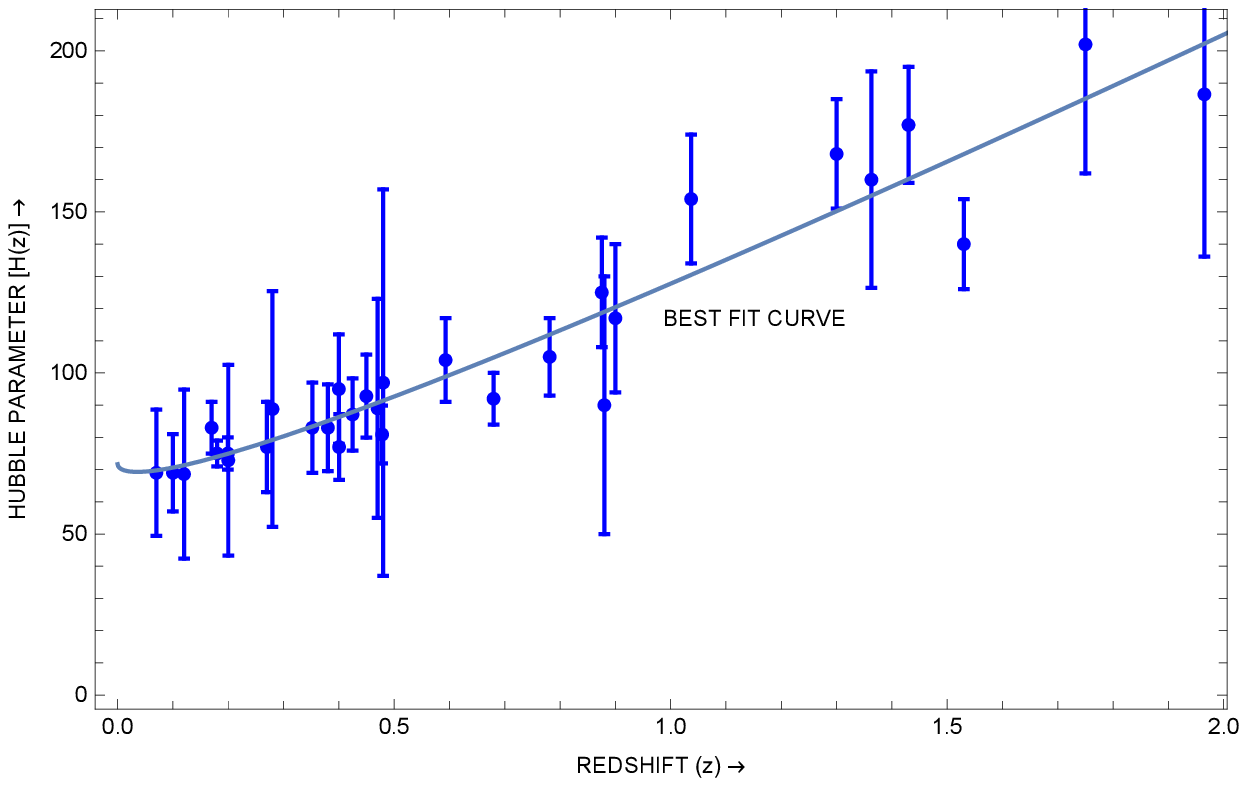}
\label{fig:subfig1} } ~~~\subfigure[   'Best fit curve' and  Hubble parameter in eq.  \eqref{eq:21b}.
 ]{
\includegraphics[width= 6.8 cm]{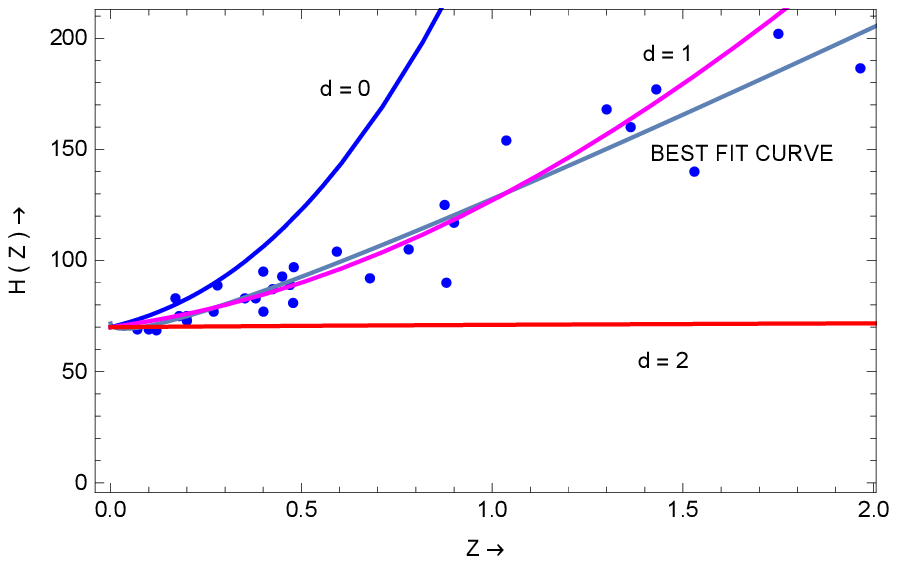}

 \label{fig:subfig2} }\label{fig:subfigureExample}~~~~~~~~~~~\caption[Optional
caption for list of figures]{\emph{ The variation of Hubble parameter $H(z)$ with Redshift  $z$  }}
\end{figure}

The apparently small uncertainty of the measurement naturally increases its weightage in
estimating $\chi^2$ statistics. We define here the   $\chi^2$ as
\begin{equation}\label{eq:20b}
\chi_{H}^2 = \sum_{i=1}^{30} \frac{[H^{obs}(z_i) - H^{th} (z_i, H_0, \theta]^2}{\sigma^2_H(z_i)}
\end{equation}
where $H^{obs}$ is the observed Hubble parameter at $z_i$ and $H^{th}$ is the corresponding
 theoretical Hubble parameter  given by eq. \eqref{eq:21b}. Also, $\sigma_H(z_i)$ denotes the uncertainty
 for the $i_{th}$ data point in the sample and $\theta$ is the model parameter.
  In this work, we have used the latest observational $H(z)$ dataset consisting of $31$
   data points in the redshift range, $0.07 \leq z \leq  1.965$, larger than the redshift
   range that is covered by the type Ia supernova. It should be noted that the confidence levels
 $1 \sigma(68.3\%)$,  $2 \sigma(95.4\%)$  and $3 \sigma(99.7\%)$ are taken proportional to
$\triangle \chi^2 = 2.3 $, $6.17$ and $11.8$ respectively, where $\triangle \chi^2 = \chi^2(\theta) - \chi^2 (\theta* )$   and  $\chi^2_m$ is the minimum value of $\chi^2$. An important quantity which is used for data fitting process is
\begin{equation}\label{eq:20c}
\overline{\chi^2} =  \frac{\chi_m^2}{dof}
\end{equation}
where subscript {\it dof}  is the  degree of freedom, and it is defined as the difference between all observational data points and the number of free parameters. If  $\frac{\chi_m^2}{dof} \leq 1$, we get a good fit and the observed data are consistent with the considered model.
\begin{figure}[ht]
\begin{center}
  \includegraphics[width=8cm]{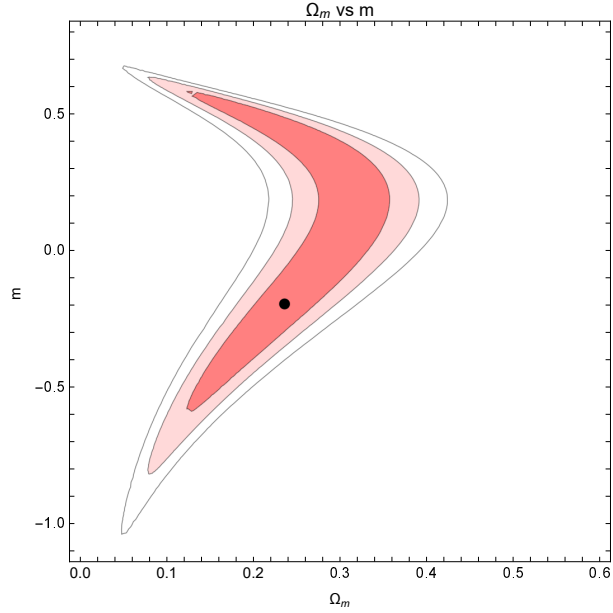}
  \caption{
  \small\emph{The contour of $\Omega_m$ vs $m$   are shown in this figure. }}
\end{center}
\end{figure}

\begin{figure}[ht]
\centering \subfigure[ Likelihood of $\Omega_{m}$  ]{
\includegraphics[width= 6 cm]{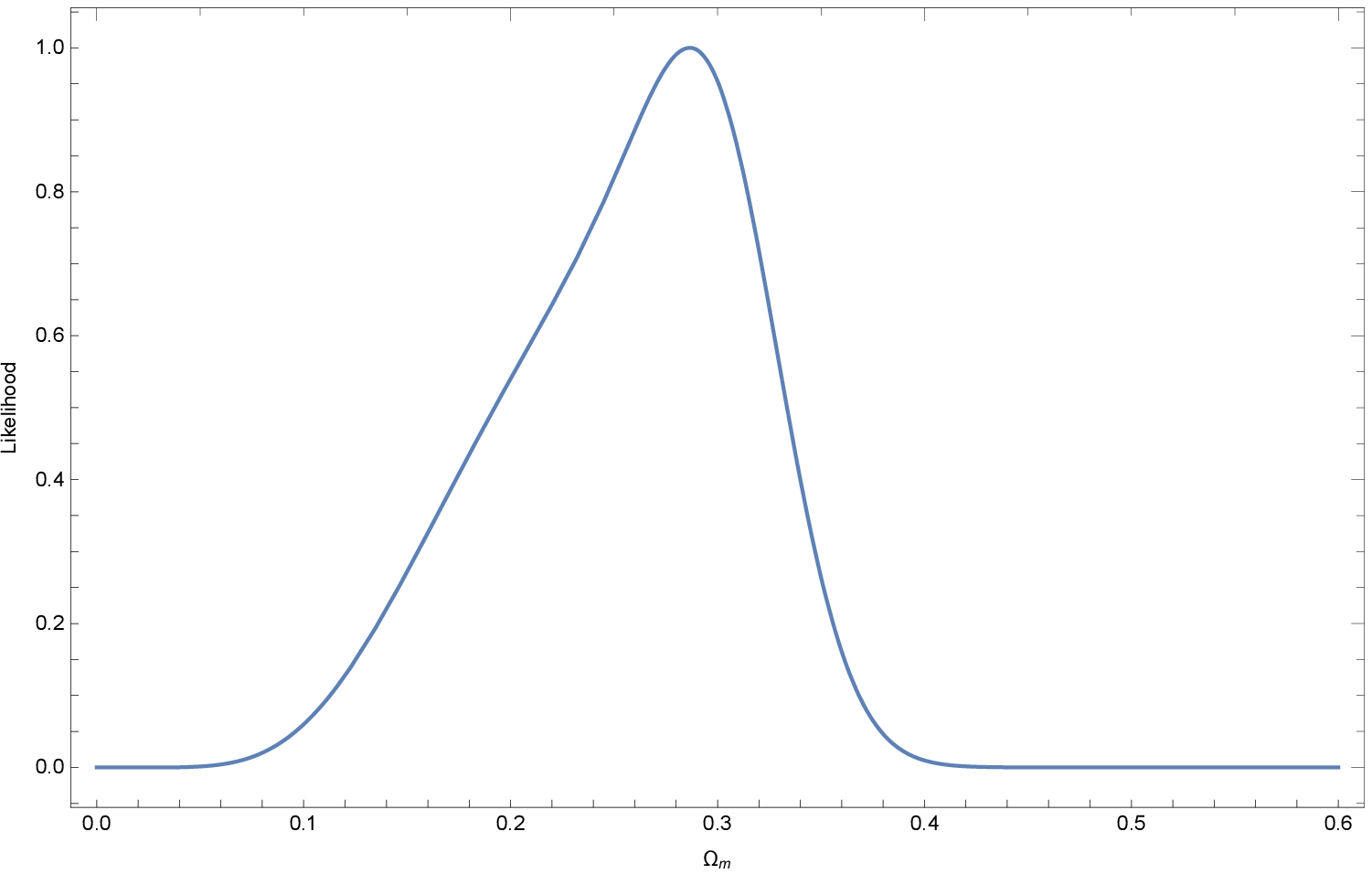}
\label{fig:subfig1} } ~~~\subfigure[   Likelihood of $m$.
 ]{
\includegraphics[width= 6 cm]{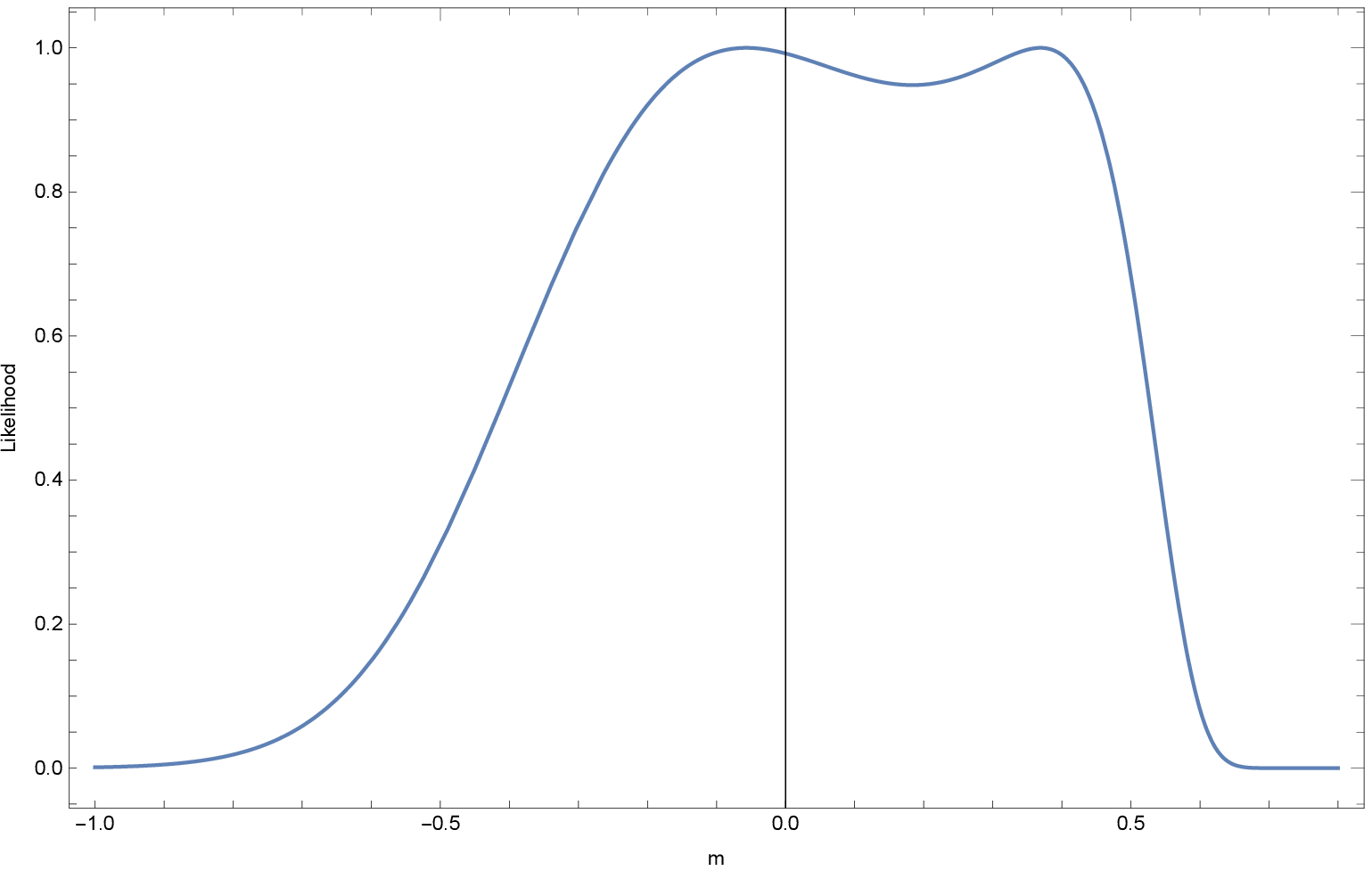}

 \label{fig:subfig2} }\label{fig:subfigureExample}~~~~~~~~~~~\caption[Optional
caption for list of figures]{\emph{ Likelihood  }}
\end{figure}

$\chi^2_{min} $  and the best-fit values of the model parameters obtained by using the SN Ia dataset and Planck-2015
results in the $1 \sigma$  confidence region is shown in Table 2.

\begin{table}[h!]
\small
  \centering
  \caption{The  $\chi^2_{min}$ value and corresponding $\Omega_m$ \& $m$. }\vspace{0.5 cm}
  \label{tab:table1}
\begin{tabular}{|c|c|c|c|}  \hline
  Parameter & $\chi^2_{min}$ & $\Omega_m$ &  m\\   \hline \hline
  Min values & 14.62 & $ 0.24 $& \ 0.44\\ \hline
  Min values & 14.62 & $ 0.24 $& \ - 0.20\\
    \hline
\end{tabular}
\end{table}
The range of $\Omega_m$ and $m$ are respectively ( 0.1257, 0.3553) and ( -0.5862, 0.5660)  in 1$\sigma$ confidence region. It is seen that the value of $m$ may be both  positive or negative. To get a dimensional reduction of extra dimensions, we consider the positive value of $m$ only. For $m = 0.54$, we use the corresponding value of $\Omega_m = 0.18$ which are lying in the 1$\sigma$ confidence region.

In a higher dimensional model with extra d-dimensions it is noted that comparing the data obtained by the differential age method (CC), the model with Chaplygin gas favours a 5D universe for a given value of $m$. It is to be mentioned here that from eq. \eqref{eq:3} in the framework of  dimensional reduction the 4-dimensional scale factor increases. But we can not explain the impact of compactification of extra dimensions  on the present acceleration of the universe.

As pointed earlier  the key eq. \eqref{eq:12} is not amenable to obtain an explicit solution  which is a function of time in known simple form. In this case the variation of cosmological variables like sale factor, flip time, dependence on extra dimensions  etc. can not be explicitly obtained. To avoid such a difficulty of obtaining solution in known form to  determine
the flip time and other physical features of cosmology we adopt here an \emph{alternative approach}  in the next section.

 \vspace{0.2 cm}

\section{ An alternative approach :}

 \vspace{0.2 cm}
In the late evolution  the universe is big enough and the second term of the right hand side (RHS) in the eq. \eqref{eq:12} is almost negligible compared to the first
term.  We know that the Chaplygin gas equation of state explains only from dust dominated era to present accelerating universe. As the 4D scale factor is  large enough it may not be inappropriate to consider only the first order
approximation of the binomial expansion of RHS of the eq. \eqref{eq:12}.
 We obtain an exact solution of the first order approximation
in the eq. \eqref{eq:12}.  Now from eq. \eqref{eq:12} we determine the late stage of evolution of the universe neglecting the higher order terms which is given by
\begin{equation}\label{eq:22}
\frac{\dot{a}^2}{a^2} =  \left(\frac{4B}{kM} \right)^{\frac{1}{2}} + \frac{cM^{\frac{1}{2}}}{ B^{\frac{1}{2}} k^{\frac{3}{2}}} \frac{1}{a^{\frac{2M}{2-dm}}}
\end{equation}

\begin{figure}[ht]
\begin{center}
  \includegraphics[width=8 cm]{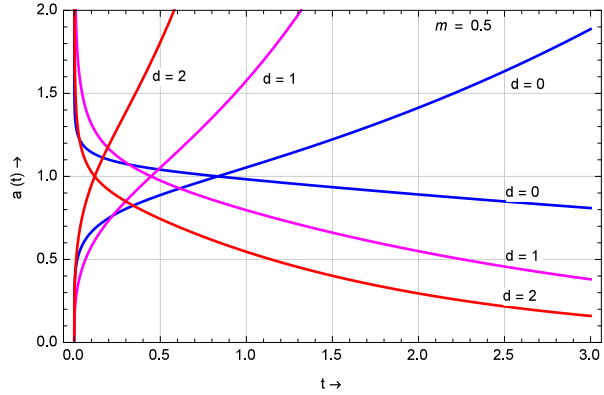}
  \caption{
  \small\emph{The variation of $a$ , $b$ with $t$  for
different values of $d$ with $B = 1.0 $ \& $c =1.0$ }\label{abt}
    }
\end{center}
\end{figure}
the eq.\eqref{eq:22} yields  as first integral an expression of the scale factor
\begin{equation}\label{eq:23}
a(t) = a_{0} \sinh^{n} \omega t
 \end{equation}
where, $a_{0} =  \left \{ \frac{cM}{2 Bk} \right \}^{\frac{2-dm}{2M}}$ ; $n = \frac{2-dm}{2M}$ and $\omega = \sqrt{\frac{2}{k}} \left(\frac{Bk}{M}\right)^{\frac{1}{4}} \frac{M}{2-dm}$.
In figure-\ref{abt},  it is evident that the evolution of the scale factor $a (t)$  and the reduction of extra dimensional scale factor $b(t)$ with time $t$ is determined by different values of $d$. This shows that the desirable feature of dimensional reduction of extra scale factor is possible. It is also seen that the rate of growth of $4D  $ scale factor depends on the number of dimensions and it is higher as the number of extra dimension increases. Again the reduction rate of extra dimensional scale factor is faster for higher dimensions. So  it is physically realistic to consider the presence of extra dimension which enhances the acceleration.
In this context, it is to be remembered that the observational results hold only for one extra dimension in our model.

Now using eqs. \eqref{eq:5}, \eqref{eq:6} and \eqref{eq:23} we can write the expression of $p$ and $\rho$ as follows.

\begin{equation}\label{eq:24}
\rho =  \left(\frac{Bk}{M} \right)^{\frac{1}{2}} \coth^2\omega t
 \end{equation}

and

\begin{equation}\label{eq:25}
p = -  \left(\frac{BM}{k} \right)^{\frac{1}{2}} \left(2- \coth^2\omega t \right)
 \end{equation}

The effective equation of state is given by

\begin{equation}\label{eq:26}
w (t) = \frac{p}{\rho} = - \frac{M}{k} \left(2 tanh^2 \omega t -1 \right)
 \end{equation}
here $w (t)$ \footnote{\begin{tiny} {By rescaling of time parameter the eq. \eqref{eq:26} may also be written as  $w(t) = -\frac{2M}{k} tanh^2 \omega (t-t_0)$, when $t = t_0$, $w(t) =0$, i.e., $p = 0$ implying dust dominated universe} \end{tiny}}
 is a function of time.

\begin{figure}[ht]
\begin{center}
  \includegraphics[width=8cm]{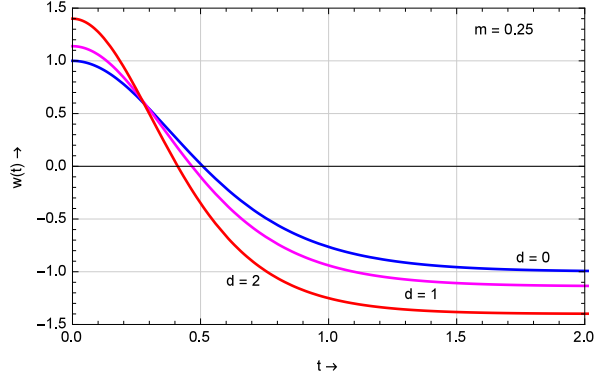}
  \caption{
  \small\emph{The variation of $w$ vs $t$  with different d are shown in this figure ( $B = 1$). }\label{wt}
    }
\end{center}
\end{figure}

From eqs. \eqref{eq:13} and \eqref{eq:26}  we get the deceleration parameter

\begin{equation}\label{eq:27}
q = \frac{1-n \cosh^2 \omega t}{n \cosh^2 \omega t}
 \end{equation}
The eq. \eqref{eq:27} gives that the exponent $n$ which determines the evolution of $q$. A numerical analysis using eq. \eqref{eq:27} shows that (i) if  $n > 1$ one gets only  acceleration, no \emph{flip} occurs in this condition. But  the eq. \eqref{eq:18a} leads to   a physically unrealistic matter field for $n > 1$. (ii) Again, if $0<n<\frac{2}{3}$ it gives early deceleration and late acceleration, so the desirable feature of \emph{flip} occurs which agrees with the observational analysis.

\begin{figure}[ht]
\begin{center}
  \includegraphics[width=8cm]{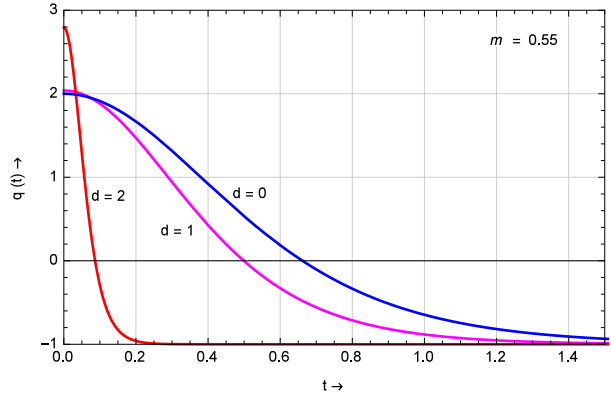}
  \caption{
  \small\emph{The variation of $q$ with $t$  for different values of d are shown in this figure ( $B=1$ ). This figure clearly shows that late flip occurs at lower dimensions.   }\label{qt}
    }
\end{center}
\end{figure}

Figure-\ref{qt}  shows the variation of $q$ with $t$ for different values of $d$ where flip occurs. It is seen that the flip occurs faster  for more dimensions.

\begin{equation}\label{eq:28}
t_f = \frac{1}{\omega} \cosh^{-1} \left(\sqrt{\frac{1}{n}} \right)
\end{equation}

 \begin{figure}[ht]
\begin{center}
  \includegraphics[width=8cm]{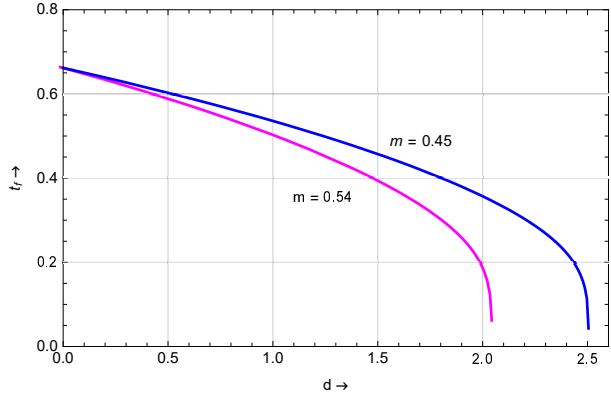}
  \caption{
  \small\emph{The graphs clearly show that flip time depends on  dimensions.   }\label{tfd}
    }
\end{center}
\end{figure}
Using eq. \eqref{eq:28} we have drawn the figure-\ref{tfd} where the variation of $t_f$ with $d$  is shown. It is seen that the flip time is lower for higher dimensions, i.e., it gives  the early acceleration for higher dimensions.

\section { Summary :}
In the paper we present a higher dimensional cosmological model to
explain the recent acceleration with a Chaplygin type
of gas.  The salient features of this model are briefly as follows:\vspace{0.5cm}

The most important thing is
that  depending on some initial conditions, the effective equation
of state during late evolution interpolates between $\Lambda CDM$ and Phantom type of expansion. In this respect our work
recovers the effective equation of state (for large scale factor)
for an analogous work of Guo \emph{et al} ~\cite{guo} where a very generalised Chaplygin
type of gas is taken. One may mention that our solutions are quite general in nature because all the well
known results of 4D Chaplygin driven cosmology are recovered when $d = 0$.

While working on any higher dimensional model one always looks for situation where dimensional reduction takes place
and the cosmology eventually becomes 4D one. Interesting to point out that our present model satisfies this important criteria for positive $m$.
It is to be noted that with the help of observational data and following $\chi^2$ minimization programme we find the range of $\Omega_m$ and $m$ are respectively ( 0.1257, 0.3553) and ( -0.5862, 0.5660)  in 1$\sigma$ confidence region. One takes the value of  $m = 0.54$ and the corresponding $\Omega_m = 0.18$ which are lying in the 1$\sigma$ confidence region.  The best fit graph is drawn from the observational data and it is seen that the graph favours only one extra dimension. That means the Chaplygin gas is apparently dominated by a 5D world.

To end the section a final remark may be in order. Being highly nonlinear one can not get a solution of the key eq. \eqref{eq:12} in a closed form forcing us to look for solutions in the asymptotic regions only. So we can not explain the evolution of 4D scale factor or reduction of extra dimensions etc. in a general way. To compensate for these incompleteness an \emph{alternative approach} is suggested where only the first order terms of the binomial expansion  are considered. By the above approach we get the time explicit solution of 4D scale  factor  $a(t) $ as well as the expression of extra dimensions $b(t)$.  It is also seen that the rate of dimensional reduction is higher for higher dimensions. So we may conclude that the effect of compactification of extra dimension  helps the acceleration.  We also investigate dimensional dependence on the deceleration parameter $q$  and flip time $t_f$. It  clearly shows that early \emph{flip} occurs for higher dimensions.

\textbf{Acknowledgment : }

DP acknowledges financial support of Sree Chaitanya College, Habra for a Minor Research project vide no SCC/MRP/2019-20/03.


\begin{thebibliography}{40}
\bibitem{pd} T. Padmanabhan  - `Understanding our Universe : Current status and open issues'  and references
therein, gr- qc/0503107
\end{thebibliography}
\end{document}